\def\K{{\rm K}}

\def\Mpc{{\rm Mpc}}

\def\km{{\rm km}}
\def\s{{\rm s}}

\def\ie{{\frenchspacing\it i.e.}}

\def\beq#1{\begin{equation}\label{#1}}
\def\eeq{\end{equation}}
\def\beqa#1{\begin{eqnarray}\label{#1}}
\def\eeqa{\end{eqnarray}}

\def\spose#1{\hbox to 0pt{#1\hss}}
\def\simlt{\mathrel{\spose{\lower 3pt\hbox{$\mathchar"218$}}
     \raise 2.0pt\hbox{$\mathchar"13C$}}}
\def\simgt{\mathrel{\spose{\lower 3pt\hbox{$\mathchar"218$}}
     \raise 2.0pt\hbox{$\mathchar"13E$}}}
\def\simpropto{\mathrel{\spose{\lower 3pt\hbox{$\mathchar"218$}}
     \raise 2.0pt\hbox{$\propto$}}}

\def\ed{\end{document}}

\def\Om{\Omega_{\rm m}}

\def\L{{\cal L}}

\def\k{{\bf k}}
\def\p{{\bf p}}
\def\d{{\bf d}}
\def\x{\hat {\bf x}}

\def\de{\delta}
\newcommand{\lexp}{\mathop{\langle}}    
\newcommand{\rexp}{\mathop{\rangle}}    
\newcommand{\rexpc}{\mathop{\rangle_c}} 

\def\fun#1#2{\lower3.6pt\vbox{\baselineskip0pt\lineskip.9pt
        \ialign{$\mathsurround=0pt#1\hfill##\hfil$\crcr#2\crcr\sim\crcr}}}
\def\la{\mathrel{\mathpalette\fun <}}
\def\ga{\mathrel{\mathpalette\fun >}}

\documentstyle[emulateapj,danonecolfloat]{article}

\begin{document}
\twocolumn[


\journalid{337}{15 January 1989}
\articleid{11}{14}

\submitted{\today. To be submitted to ApJ.}



\title{Inferring the Linear Power Spectrum from the
Lyman-$\alpha$ Forest}

\author{Matias Zaldarriaga$^{1}$, Rom\'an Scoccimarro$^{1}$, \& Lam 
Hui$^{2}$}

\affil{$^{1}$Physics Department, New York University, 4 Washington
Place, New York, NY 10003}

\affil{$^{2}$Department of Physics, Columbia University, 538 West
120th Street, New York, NY 10027}

\vskip 1pc

\begin{abstract}

We discuss the determination of the linear power spectrum of dark
matter fluctuations from measurements of the transmitted flux
Ly-$\alpha$ Forest power spectrum.  We show that on most scales probed
by current measurements, the flux power spectrum is significantly
affected by non-linear corrections to the linear dark matter power
spectrum due to gravitational clustering.  Inferring the linear dark
matter power spectrum shape is therefore difficult due to non-linear
corrections driving the non-linear power spectrum to a $k^{-1.4}$
shape nearly independent of initial conditions.  We argue that some
methods used in previous estimates underestimate the uncertainties in
the shape of the linear dark matter power spectrum.

\end{abstract}


]


\section{Introduction}
\label{introduction}

In the last few years the Lyman $\alpha$ forest has become
increasingly important as a probe for cosmology.  The availability of
high resolution high signal-to-noise spectra, mainly from the HIRES
spectrograph on the Keck telescope, has revolutionized the field.  The
data have been compared with the theoretical predictions of currently
favored cosmological models, mainly using numerical simulations
(\cite{chen94,her95,zhang95,m96,mucket96,wb96,th98}).  The success of
these models has been impressive.  The observed properties of the
forest are very well reproduced even though the models were primarily
constructed in the context of galaxy clustering -- this was a
non-trivial accomplishment.  Examples of comparisons of model
predictions with observational data can be found in
Croft et al. (1998), Bryan et al. (1999), McDonald et al. (1999, 2000),
Nusser \& Haehnelt (2000), Schaye et al. (1999, 2000), 
Croft et al. 2000, Meiksin, Bryan \& Machacek (2001), 
Choudhury, Srianand \& Padmanabhan (2001), 
Zaldarriaga, Hui \& Tegmark (2000) (ZHT2000), and
Pichon et al. (2001).  The same cosmological
models also produce acceptable fits to many other cosmological
observations such as the power spectra of CMB fluctuations, the
luminosity distance to high redshift supernovae, the number density of
massive clusters, etc.  (see \cite{wtz01} for a recent summary).

The success of the numerical models of the forest has also led to the
construction of several analytic approximations (e.g.
\cite{BD97,GH96,Hui97b,viel01,joop01}). 
The physics that is responsible for the absorption seems to be well
understood so the hope has emerged that we could use the
Lyman-$\alpha$ forest to further constrain cosmological models.  The
advantages of the forest over other observations are clear, since we
get to observe the universe at an earlier time we may be able to infer
the rate of growth of density fluctuations.  It is also assumed that
because we are observing at a higher redshift, scales that today (e.g.
in galaxy surveys) are in the ``non-linear regime'', can be probed
when they are still ``linear'', thus greatly simplifying the analysis. 
In this paper, we challenge this assumption.

Studies of the forest have other applications.  Through the
Alcock-Paczynski test (Alcock \& Paczynski 1979) applied to the
correlation functions for pairs of lines of sight, it seems possible
to directly infer the geometry of the universe (Hui, Stebbins \& Burles 2000, 
McDonald \& Miralda-Escude 2000, McDonald 2001).  Higher order statistics of the
flux may also prove useful in testing that the structure observed in
the flux is actually produced by gravitational instability
(Zaldarriaga, Seljak \& Hui 2001) and in testing if there are
fluctuations in the temperature of the IGM, left for example as
remnants of a late ionization of Helium II (Zaldarriaga 2001, Theuns
et al. 2001).

The power spectrum of the flux is perhaps the most studied of the
statistics of the Lyman $\alpha$ forest flux.  Inferring cosmological
parameters from the power spectrum is however a difficult challenge. 
A naive approach would be to run hydrodynamical simulation for enough
models to cover the parameter space of interest and compare the
results with the observed data.  This approach is impractical
as the computer time needed to run the number of simulations required
is enormous.  One should also remember that on top of the usual
cosmological parameters that are needed to specify the cosmological
model ({\ie} the energy densities in the different components and the
power spectrum of initial density fluctuations) in the case of the
forest one needs to specify the time evolution of the background of
ionizing radiation that is responsible for ionizing and heating the
gas.  This requires the introduction of additional parameters.  One
should also keep in mind that even if we were able to run a large
enough grid of hydrodynamical simulations, it is not obvious that all
the relevant physics is correctly included in these simulations,
``feedback'' effects from the forming stars and galaxies being the
obvious example (see e.g. \cite{mbm01} for possible conflicts with
observations). 

A significant simplification can be achieved with the introduction of
the N-body model of the forest.  In this model the spatial
distribution of the dark matter is obtained by running an N-body
simulation.  The gas is then assumed to trace the dark matter (or
maybe a smoothed version of the dark matter field to account for the
effect of gas pressure).  Furthermore it is assumed that the gas is in
photo-ionization equilibrium and follows a power-law equation of
state.  This simple prescriptions lead to a simulated
Lyman $\alpha$ forest that is in reasonable agreement with the
predictions of full hydro simulations.  As could have been expected,
the agreement is good but not perfect (\cite{mw00}).  The great advantage of
the N-body approximation is that it significantly reduces the
computational burden of exploring parameter space.  A first attempt at
implementing the straightforward method of running a grid of N-body
simulations and comparing that with the forest data to extract
parameters was presented in ZHT2000.

In the quest of incorporating data from the forest to studies of
cosmological parameters even the N-body model is computationally too
expensive.  This has lead to the introduction of approximate
techniques to invert the observed flux power spectrum to get the
3D {\it linear} matter power spectrum of the dark matter.  The results
of these inversions are then compared to any model of interest and
the results of these comparisons are used to constrain parameters.  It
is clear that the inversion method greatly simplifies the
comparison between theory and observations, which only requires the ability
to predict the {\em linear} power spectrum of any given model, a
straightforward task.  Obviously the relation between the linear dark matter
power spectrum and the flux power spectrum is non-trivial, so these
inversions are not rigorous inversions in the mathematical sense.  The
resulting prescriptions, however, are calibrated with numerical
simulations, and are the result of detailed studies of the physics of 
the forest (Croft et al. 2000). 

The aim of this paper is to point out the limitations of inversion
techniques which attempt to match predictions of the forest from
knowledge of the linear power spectrum of dark matter.  In particular,
we point out that the evolution of fluctuations at $z \sim 3$ is
significantly non-linear at scales that can be orders of magnitude
larger than estimated by naive linear arguments.  Furthermore,
non-linear corrections tend to drive the density (and velocity) field
to ``universal'' spectra nearly independent of initial conditions,
erasing information and leading to weak constraints on the shape of
the primordial dark matter power spectrum. 

The rest of this paper is organized as follows.  In \S \ref{Nbody} we
summarize the constraints on the primordial power spectrum that we
obtain by running a grid of N-body simulations 
and compare them with those obtained
by the inversion technique described above.  
The remaining sections present what
we believe to be the main reasons behind these results.  In \S
\ref{nonlinear} we discuss the role of non-linear corrections, in \S
\ref{simple} we discuss the effects of the transformation between
density and flux and the importance of redshift distortions.  We
conclude in \S \ref{conclusions}.

\section{Likelihood from N-body simulations}
\label{Nbody}

Our first objective is to convince the reader that there are problems
with the error bars that result from inversion methods.  We use the
technique described in ZHT2000 to find constraints on the amplitude
and slope of the linear power spectrum of dark matter fluctuations. 
Basically we run a grid of P$^{3}$M N-body simulations of the standard
CDM model in boxes of size $L=16\ h^{-1} \Mpc$.  We choose a range of
spectral indices for the primordial power spectrum, ($n_p=0.4,0.7,
0.9, 1.0, 1.1, 1.3, 1.7$), $n_{p}=1$ corresponding to the usual
scale-invariant spectrum, and store the outputs at several different
times, corresponding to scale factors $a=(.05, .065, .085, .11, .14,
.19, .24, .31, .41)$.

The relation between gas density and dark matter density is modeled by
introducing additional parameters.  To model the effects of pressure
we smooth the density field with a three-dimensional smoothing scale
we call $k_f$. For simplicity we used a Gaussian window,
$W(k)=\exp(-(k/k_f)^2$ as described in ZHT2000.
The smoothing scale $k_f$ is related to the Jeans
scale $k_J$ by, 
\beq{kf1} k_f = \eta k_J; \ \ k_J = a c_s^{-1} \sqrt{4\pi G \bar
\rho}, 
\eeq 
where $c_s$ is the sound speed of the gas and $\eta$ is a
function of time and the details of the ionization history with values
around $\eta \sim 2$ for standard assumptions (\cite{gh98}). 
Equation (\ref{kf1}) leads to, 
\beqa{kf2} k_f\approx 30 \ h\Mpc^{-1}\ \times \Big({\eta \over 2}\Big) \Big({h
\over 0.7 }\Big)^{-1} \nonumber \\ \times 
\Big({\Om h^2\over 0.15 }\Big)^{1/2} \Big({1+z\over 4}\Big)^{1/2} 
\Big({ T\over 10^4 \K}\Big)^{1/2} \Big({ k T\over
c_s^2 m_p}\Big)^{1/2},
\eeqa
where $m_p$ is the mass of the proton. 

The equation of state of the gas is described by two parameters, the
temperature at zero overdensity $T_0$ and the slope $\alpha$,
$T=T_0(1+\delta)^{\alpha}$ where $\delta$ is the overdensity (Hui \& Gnedin 1997a).  The
optical depth of a given fluid element is given by $\tau=A
(1+\delta)^{\beta}$, the assumption of ionization equilibrium leads to
$\beta=2-0.7\alpha$.  The amplitude $A$, which physically depends on
the strength of the ionizing background, is chosen so that the
constructed spectra have the observed mean transmission which we call
$\bar F$.  For each simulation we construct grids of model predictions
with $(k_f,T_0,\alpha,\bar F)$ as parameters.  Our full parameter
vector $\p=(a,n_p,k_f,T_0,\alpha,\bar F)$ has 6 dimensions.  

For each model in our six dimensional space 
we calculate the predicted flux power spectrum $P_{f}(\p)$. We
compare these prediction to the data and compute a likelihood function as
described in ZHT2000. We used a Gaussian
approximation for the likelihood, 
\beq{chi2eq} 
\L(\d;\p)\propto  \prod_i
\exp\left[-{1\over 2}\left({d_i-P_{fi}(\p) \over
\sigma_i}\right)^2\right], 
\label{like1}
\eeq 
where $i$ runs over the different data points $d_i$, the model
prediction for that wavevector are $P_{fi}(\p)$ and $\sigma_i$ are the
error bars on each point. To obtain one dimensional constraints on individual
parameters or two dimensional constraints for a pair of parameters, 
we marginalized the likelihood one parameter at a time,
reducing the dimensions of the grid by one in each step until only the
parameter or pair of parameters of interest was left.   

When marginalizing over parameters we allow $\bar F$ to vary within
observational constraints. At $z=2.72$ corresponding to the mean
redshift of the sample used we take $\bar F=0.707\pm 0.035$, a $ 5 \%$
uncertainty. We should note that we have added in quadrature a
``theoretical uncertainty'' to the observational errors. This extra
uncertainty was estimated from the fluctuations in the mean flux
computed in different realizations of the same model and could be
improved upon with better simulations.  We include the constraints on
$\bar F$ by multiplying the likelihood coming from the forest power
spectrum by $\exp(-\delta\chi^2/2)$, with $\delta\chi^2=[(\bar
F-0.707)/0.035]^2$.

In ZHT2000 we presented
constraints on the primordial power spectrum. In that paper
we were mainly interested in the effects of the temperature and the
constraints that could be obtained on it.  The temperature smooths
the spectra and we wanted to make sure that we did not underestimate
the uncertainties by assuming we knew what $k_f$, the other source of
smoothing, actually was.  For that reason we had a grid in $k_f$ that
was too conservative, $k_f=(5,10,20,30,40,45,50,55,60,70,80) h\
\Mpc^{-1}$.  The smallest values in this grid (the largest smoothings)
produce much more smoothing than can be expected from pressure. 
However because $k_f$ smooths the field in 3D before we apply the
non-linear transformation that takes density to optical depth and
flux, while $T_0$ acts in 1D smoothing the optical depth, we were able
to distinguish between them.  We also found that the $k_f$ smoothing
was quite degenerate with the spectral index of the perturbations $n$
so our constraints on that parameters were weak.

In this paper we improve on the earlier analysis by using newer data
from Croft et al. (2000).  We use the power spectra of their ``fiducial
sample'' which has mean redshift $z=2.72$ (table 3 of \cite{croft00}).
The sample was compiled using data from the Low Resolution Imaging
Spectrometer (LRIS) and the High Resolution Spectrometer (HIRES) on the Keck
Telescope. Data in the redshift range $2.3<z<2.9$ with a total
path length $\delta z \approx 25$ was used to create the ``fiducial
sample'' (\cite{croft00}).  As another improvement in our
marginalizations we restrict the $k_f$ grid to values $k_f > 30 h\
\Mpc^{-1}$.  With the improved data and a more physical ``prior'' for
$k_f$ we hope to find tighter constraints on the primordial power spectrum.

\begin{figure}[tb]
\centerline{\epsfxsize=9cm\epsffile{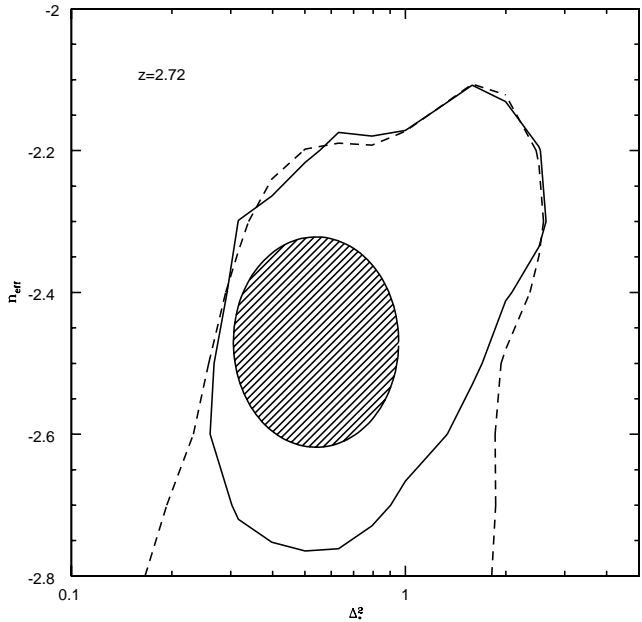}}
\caption{Constraints in the $\Delta-n_{\rm eff}$ plane.  The solid
line shows constraints when we fix $k_f=30 h\ \Mpc^{-1}$, the dashed
curve is when we allow $k_f > 30 h\ \Mpc^{-1}$.  In both cases the
lines correspond to an increase of $\delta \chi^2 = 6.17$ with respect
to the best model ($95\%$ confidence for a Gaussian).  The dashed
circle are constraints obtained using the inversion technique and
the same data by \cite{croft00}.}
\label{sgneff}
\end{figure}

In figure \ref{sgneff} we show our constraints in the $\Delta_*^2-
n_{\rm eff}$, the amplitude and slope of the linear power at a scale $k_p=0.03 \ (\km /
\s)^{-1}$. That is $\Delta^2(k)=4\pi k^3 P(k)$ and $\Delta^2(k_p)= 
\Delta_*^2$ with $(d\ln \Delta^2(k)/d\ln k)|_{k_p}=n_{eff}+3$. 
To illustrate the effect of the uncertainty in $k_f$ we show the results if we
marginalize over $k_f$ with the restriction $k_f > 30 h\ \Mpc^{-1}$ or
if we just assume $k_f=30 h\ \Mpc^{-1}$.  For comparison we also plot
the results obtained by \cite{croft00} with the inversion technique. 
Note that when we marginalize over $k_f$, that is when we allow
smaller $k_f$ smoothing scales, models with smaller spectral indices
become allowed.  This makes sense because one can compensate for the
lack of power in the dark matter by assuming that the smoothing due to
pressure is smaller.

\begin{figure}[tb]
\centerline{\epsfxsize=9cm\epsffile{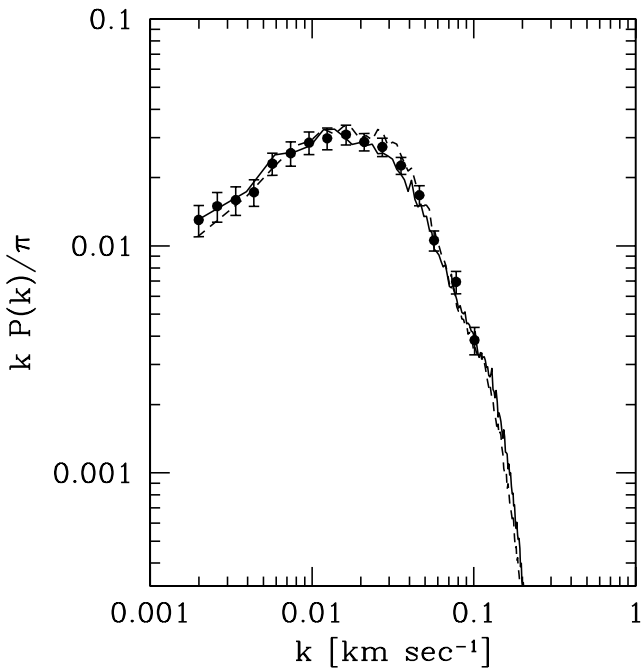}}
\caption{Transmission power spectrum from Croft et al.  (2001)
together with two models that are acceptable fits.  The two models
have parameter vectors $\p=(0.19,0.7,30,250,0.1,0.684)$ and
$\p=(0.11,1.1,30,250,0.3,0.684)$.}
\label{psfig}
\end{figure}

Figure \ref{sgneff} shows that we find acceptable models outside the
range inferred from the inversion technique.  In figure \ref{psfig} we
show examples of such models.  The two models have very different
spectral indices of the linear power spectrum, one has $n_{p}=0.7\ \
(n_{\rm eff}=-2.7,\Delta^2_*=0.42)$ and the other has $n_{p}=1.3\ \
(n_{\rm eff}=-2.3,\Delta^2_*=0.57)$.  Part of the increase in the
allowed range of $\Delta^2_*$ seen in figure \ref{sgneff} comes from
our more conservative choice of errors for $\bar F$. Unfortunately,
the relation between the amplitudes of the dark matter and flux power
spectrum is very sensitive to $\bar F$, a quantity that is sensitive
to small scale physics and may not be very well modeled by dark
matter only simulations.  Figures \ref{sgneff} and \ref{psfig} clearly
show that the inversion technique underestimates the errors, thus
results from such studies should be interpreted with care if one wants
to incorporate them to likelihood analysis of cosmological parameters.

It is important to understand that the non-linear corrections make the
power spectra of the two models under consideration look similar but
that in order to fit the data we must allow some extra freedom in
another parameter to accommodate the residual differences.  In the
example shown, $\alpha$ had to vary between $0.1$ and $0.3$. This
means that if there are independent ways (not using the power
spectrum) to measure $\alpha$, and in general the equation of state,
then the error-bar on $n$ can conceivably be reduced. Using the
b-parameter-column density distribution might be one way of achieving
this, but the error on $\alpha$ is at present still quite large
(eg. Ricotti et al. 2000, Schaye et al. 2000, McDonald et al. 1999,
Bryan et al. 1999, Hui \& Rutledge 1999).

We should emphasize that the error-bars we find using our grid method
are likely to underestimate the real uncertainties because we do not
include uncertainties associated with hydrodynamic effects.
Nonetheless, it is possible that our uncertainties are larger than
they should be.  For example, the smoothing due to pressure ($k_f$),
the temperature of the gas and the slope of the equation of state can
all be determined once the ionization history is known [i.e. equations
(\ref{kf1}) and (\ref{kf2}) for $k_f$].  However we are allowing them
to vary independently in our marginalization.  This means that if we
were to factor in some ``physical priors'' that restricted these
parameters or forced relations between them, our constraints could get
tighter.

The aim of the following sections is to try to identify what is
causing the increased uncertainties.



\section{Non-linear Corrections: Critical Index}
\label{nonlinear}

\begin{figure}[tb]
\centerline{\epsfxsize=9cm\epsffile{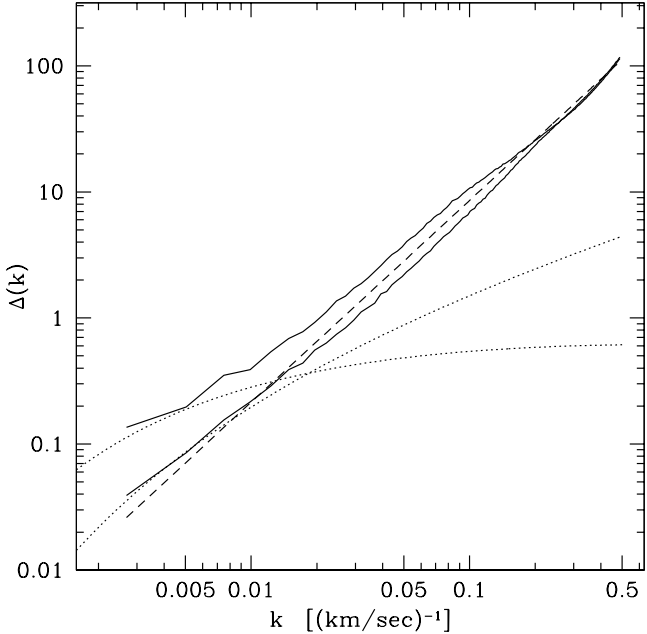}}
\caption{Linear (dotted) and non-linear (solid) dark matter power
spectra for two models with primordial spectral index $n_{p}=0.7$
(top) and $n_{p}=1.3$ bottom, corresponding to the transmission power
spectra shown in figure~\ref{psfig}.  Note how non-linear corrections
are very large for the $n_{p}=0.7$ case, even though the linear
spectrum suggests linear PT should be accurate.  Non-linear
dynamics tends to establish a power spectrum with a critical index
$n=-1.4$, as shown by the dashed line.}
\label{pkfig}
\end{figure}

The scales probed by the Lyman-$\alpha$ forest at $z \sim 3$
correspond to values of the {\em linear} mass power spectrum $P_{L}(k)$
where $4\pi k^{3} P_{L}(k) \sim 0.01-1$.  This may suggest that these
scales are safely in the linear regime.  However, the amplitude of
linear fluctuations is not the only quantity that determines the
magnitude of non-linear corrections.  The other important quantity is
the spectral index at scales in the weakly non-linear regime (Makino,
Sasaki \& Suto 1992; {\L}okas et al.  1996; Scoccimarro \& Frieman
1996).  For typical CDM models, the effective spectral index $n_{\rm
eff}$ at the non-linear scale ($k_{\rm nl}$ such that $4\pi k_{\rm
nl}^{3} P_{L}(k_{\rm nl}) \equiv 1$) is quite negative $n_{\rm eff}
\la -2.5$ (e.g. compared to $z=0$ where $n_{\rm eff} \sim -1.5$). 
This means that non-linear corrections are large, since for steep
spectra gravitational collapse is driven by large-scale coherent flows
which substantially enhance the growth of density perturbations
compared to the linear case.  In fact, for power-law initial spectra
we can write the non-linear mass power spectrum in one-loop perturbation
theory (PT) as (Scoccimarro \& Frieman 1996)

\begin{equation}
    P(k) = P_{L}(k) \Big[ 1 + \alpha(n) \Big( \frac{k}{k_{\rm nl}} 
    \Big)^{n+3} \Big],
    \label{eq:P1L}
\end{equation}

where $\alpha(n)$ is a function of spectral index which monotonically
decreases with $n$, becoming larger than unity for $n<-1.7$.  For $n
<-1.4$, the non-linear correction is positive, leading to a non-linear
spectrum less steep than the linear one.  For $n \approx -1.4$,
$\alpha$ becomes zero and thus the power spectrum retains its linear
shape.  For $n>-1.4$, the non-linear corrections are actually
negative, so the non-linear power spectrum grows slower than in linear
PT, becoming steeper than the linear spectrum.  As a result of this,
the non-linear power spectrum is driven towards a ``critical index''
$n_{c}=-1.4$, regardless of its linear slope (Scoccimarro \& Frieman
1996).  Although Eq.~(\ref{eq:P1L}) is valid rigorously only for a
power-law spectrum, CDM spectra behave in the same way, as seen in
Fig.~\ref{pkfig}.  Note how the shape of the non-linear power spectrum
becomes $\sim k^{-1.4}$ (dashed line) even at $k < k_{\rm nl}$.  In
fact all the scales shown in the figure are below $k_{\rm nl}$ for the
model with $n_p=0.7$.  Similar behavior is seen in Fig.~1 of Narayanan
et al.  (2000) and Fig.~3 of White \& Croft (2000) for CDM models with
truncated linear spectra.

\begin{figure}[tb]
\centerline{\epsfxsize=9cm\epsffile{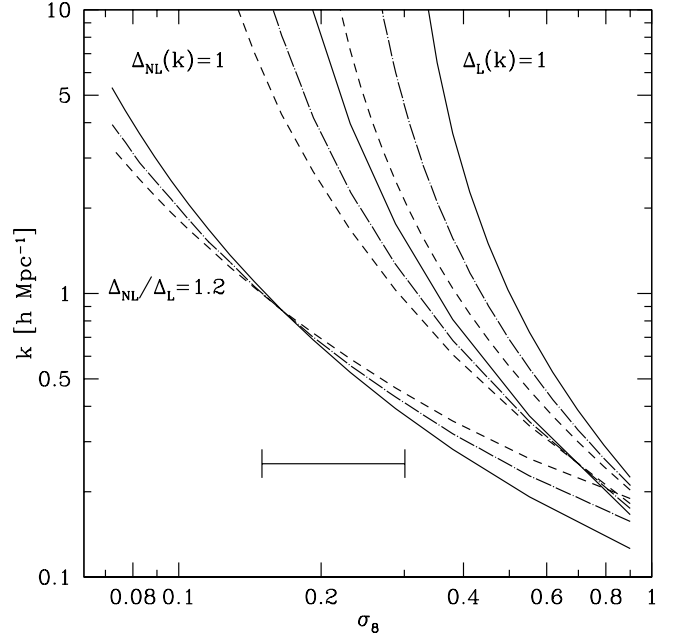}}
\caption{The importance of non-linear effects, as a function of
$\sigma_{8}$ for linear power spectra with shape parameter
$\Gamma=0.21$ (dashed), $\Gamma=0.15$ (dot-dashed) and $\Gamma=0.1$
(solid).  The three lines denote the non-linear scale according to
linear spectrum (top), non-linear spectrum (middle), and the scale at
which non-linear corrections are 20\%.  At low redshift, high
$\sigma_{8}$, all these scales coincide, approximately.  At high
redshift, small $\sigma_{8}$, non-linear corrections are much larger
and thus become important at wavenumbers that can be a very small
fraction of the non-linear scale.  The horizontal bar denotes
approximately the range of normalizations for $z \sim 3$.  Note that
translation from h/Mpc to inverse velocity units is given by $k_{v} =
k/(aH) \approx k/H_{0} [\Omega_{m}^{0} (1+z)]^{-1/2}$ at high $z$,
where $\Omega_{m}^{0}=\Omega_{m}(z=0)$.  For $\Lambda$CDM with
$\Omega_{m}^{0}=0.3$, $\Omega_{\Lambda}^{0}=0.7$, $k_{v}=k/111.5 $ at
$z=3$.}
\label{k20fig}
\end{figure}

Figure~\ref{k20fig} illustrates the importance of non-linear
corrections as a function of power spectrum normalization
$\sigma_{8}$, for linear power spectra with shape parameter
$\Gamma=0.21$ (dashed), $\Gamma=0.15$ (dot-dashed) and $\Gamma=0.1$
(solid), which approximately span the range of spectral indices
obtained in figure~\ref{sgneff}.  The calculation was done using
one-loop perturbation theory\footnote{If anything, this calculation
underestimates non-linear effects.}.  The scales where linear (top
three lines) and non-linear (middle lines) fluctuations are of order
unity and where non-linear corrections are 20\% (bottom lines) all
agree at large $\sigma_{8}$, where the spectral index at the
non-linear scale is about $n_{\rm eff} \sim -1.5$.  At low
$\sigma_{8}$, the effective spectral index at weakly non-linear scales
becomes much more negative, leading to stronger non-linear
corrections.  The wavenumber at which non-linear corrections are 20\%
can thus be a very small fraction of the non-linear scale (estimated
either from the linear or the non-linear power spectrum).  For the
normalizations corresponding to $z \sim 3$ (denoted by the horizontal
bar in figure~\ref{k20fig}), the wavenumber at which non-linear
corrections become important can be orders of magnitude smaller than
that estimated from linear perturbation theory.  In particular, note
that models with more negative spectral index (lower $\Gamma$) require
higher $\sigma_{8}$ to match the fluctuation amplitude at the scales
probed by the Ly-$\alpha$ forest, thus these models are most strongly
affected by non-linear effects, at wavenumbers as small as $k \sim
0.004$ km/sec.

The outcome of this situation is that at $z \sim 3$ non-linear
corrections are important at scales of interest for the Ly-$\alpha$
forest and drive the mass power spectrum to a $k^{-1.4}$ dependence. 
Therefore, measuring the Ly-$\alpha$ flux power spectrum at these
scales does not give a direct probe of the initial, linear, shape of
the dark matter power spectrum, leading to degeneracies in the range
of initial shapes constrained by the data, as shown in
figure~\ref{sgneff}.  This is illustrated rather clearly by comparing
figures~\ref{psfig} and~\ref{pkfig}, where significantly different
linear spectra display very similar transmission power spectra. 
However, before we can establish that non-linear dynamics is the main
reason behind the observed behavior, we must consider other non-linear
effects that potentially enter: the non-linear mapping between the
dark matter density field and Ly-$\alpha$ transmission, and the
effects of redshift distortions.

\section{Flux Correlations: A Simple Model}
\label{simple}

The discussion in the previous section was in terms of the dark matter
density field $\delta$.  However, as we show below, there is a
relevant range of scales for which the picture described above applies
as well to the transmitted Ly-$\alpha$ flux power spectrum.  The
reason is that the characteristic wavenumber for which the non-linear
nature of the mapping between $\delta$ and $F$ enters, the smoothing
scale wavenumber\footnote{The sharp, exponential decline in the flux
power spectrum seen in figure \ref{psfig} is due to thermal
broadening.  The scale $k_{\rm SM}$ is related to the thermal
broadening scale $k_{TH} \approx 0.1 (km/sec)^{-1}$ by the non-linear
mapping from optical depth to transmission power spectrum; this
follows from Eq.~(\ref{eq:xiF}) below.  That is, for a given value of
$\bar{F}$ and $\sigma^{2}$, $\xi_{F}/\sigma_{F}$ is a function of
$\xi/\sigma$.  Then, smoothing in the optical depth (approximately
density) can be translated into smoothing in the transmission power
spectrum.  For $\sigma \ga 1$, $k_{SM}< k_{TH}$.} $k_{\rm SM} \sim
0.05$ (km/sec)$^{-1}$, is much larger than the wavenumber where
non-linear corrections in the density field become important (see
figures~\ref{pkfig}-\ref{k20fig}).

We take a simple toy model to illustrate our arguments.  The
transmitted flux $F=\exp (-\tau)$ is approximated as $F=\exp [-A
(1+\delta)^{2}]$ and $\delta$ is the density field assumed to be
Gaussian for the moment (we will relax this assumption below).  In
this case, the mean transmission is given by

\begin{equation}
    \bar F  = \frac{1}{\sqrt{1+2A \sigma^{2}}}
    \exp \Big[ \frac{-2A}{1+2A\sigma^{2}} 
    \Big]
    \label{eq:meanF}
\end{equation}

We work in terms of the flux contrast, $\de_{F} \equiv F/\bar F
-1$, then the two-point function of $\de_{F}$ reads,

\begin{eqnarray}
    1+\lexp \de_{F}(1) \de_{F}(2) \rexp &\equiv& 
    1+\xi_{F} \nonumber \\ &=&
    \frac{(1+2A\sigma^{2})}{\sqrt{1+4A\sigma^{2}+4A^{2}(\sigma^{4}-\xi^{2})}}
     \\ & & \times 
    \exp \Big[ \frac{4A^{2} \xi}{(1+2A \sigma^{2})[1+2A 
    (\sigma^{2}+\xi)]} \Big] \nonumber 
        \label{eq:xiF},
\end{eqnarray}
where $\xi$ is the two-point function of the density field.  Note that
in the limit where $\xi \ll \sigma^{2}$ (but {\em not} necessarily
$\xi <1$, since $A \sigma^{2}$ cannot be large due to the mean flux in
Eq.~(\ref{eq:meanF}) being of order unity), the above expression
reduces to,

\begin{equation}
    \xi_{F} = \frac{4A^{2}}{(1+2A \sigma^{2})^{2}}\ \xi\equiv b^2 \xi
    \label{eq:xiF2}
\end{equation}

Thus, in this case the flux correlation function is proportional to
the density correlation function with some bias factor.  This is
certainly expected at large scales, since the flux is a local
transformation of the density contrast, it follows from well-known
arguments (e.g. Fry \& Gazta\~naga 1993, Scherrer \& Weinberg 1998)
that the correlation function of the flux should be proportional to
that of the underlying density field.  The bias factor in this simple
model, Eq.~(\ref{eq:xiF2}), can be shown to change by a factor of
about two when the the mean flux, Eq.~(\ref{eq:meanF}), changes within
its 2-$\sigma$ interval, similar to the variations seen in Fig.~17 of
Croft et al.  (2000).  Equivalently figure 10c of McDonald (2001)
shows that a 3\% change in the mean flux can lead to a 40\% change in
the power spectrum amplitude.

As pointed out above, the validity of Eq.~(\ref{eq:xiF2}) is not
restricted to the linear regime, since there are other small
parameters that enter.  The value of $A \sigma^{2}$ is controlled by
the mean flux as given by Eq.~(\ref{eq:meanF}).  Since the mean flux
has to be between 0 and 1, in particular $\bar F \approx 0.7$, $A
\sigma^{2}$ is constrained to be smaller than unity\footnote{This can
break down for non-Gaussian distributions, since as $\sigma$ increases
the PDF becomes more peaked about $\delta = -1$ and one can thus
tolerate a larger $A$.  For the models discussed in Section~2, $A
\sigma^{2} \approx 1$.}, even in the limit that $\sigma^{2} \gg 1$. 
Eq.~(\ref{eq:xiF2}) then follows as long as $\xi \ll \sigma^{2}$,
irrespective of the value of $\xi$ (since $A \sigma^{2} \la 1 $, $A
\xi \ll 1$ in this regime).  The condition that $\xi \ll \sigma^{2}$
means in the Ly-$\alpha$ case that the scales under consideration be
much larger than the thermal broadening scale, $k \ll k_{\rm TH}$, the
characteristic smoothing scale where $\sigma^{2}$ is evaluated.  This
is in any case where most of the information can be extracted, so the
approximations leading to Eq.~(\ref{eq:xiF2}) are physically relevant.

Of course, when $\xi$ approaches unity, the Gaussian model for the
density field PDF is not a good approximation anymore.  In order to
see how the arguments above generalize to the non-Gaussian case, let's
consider the expansion of the two-point PDF when $\xi \ll \sigma^{2}$,

\begin{equation}
    P(\de_{1},\de_{2}) \approx P(\de_{1}) P(\de_{2})\ [1+ b(\de_{1}) 
    b(\de_{2})\ \xi],
    \label{eq:Pfac}
\end{equation}
where $b(\de)=\de/\sigma^{2}$ in the Gaussian case.  Although the
relation in Eq.~(\ref{eq:Pfac}) was written starting from a Gaussian
PDF, its validity is much more general.  It can be obtained rigorously
in the weakly non-linear regime (in tree-level PT, Bernardeau 1996),
and in the highly non-linear regime for factorizable hierarchical
models (Bernardeau \& Schaeffer 1999), where specific predictions for
the function $b(\de)$ can be given in both cases.  Then, the linear
relation in Eq.~(\ref{eq:xiF2}) between $\xi_{F}$ and $\xi$ continues
to hold, but with a different bias factor than in the Gaussian case. 
Note that in the hierarchical model, $\xi$ is understood to be the
fully non-linear correlation function.

These two examples, however, are not exactly relevant for structure
formation at $z \sim 3$ at the scales of interest.  Tree-level PT is
not a good approximation because the spectral index at the non-linear
scale is so negative $n_{\rm eff} \la -2.5$ that loop corrections to
moments of the density field are very important, as noted above for
the power spectrum.  On the other hand, we cannot assume a
factorizable hierarchical model, which is expected to become a good
approximation only in the highly non-linear regime.

However, the structure of loop corrections still respects
Eq.~(\ref{eq:Pfac}) in the following broad sense.  Consider the
calculation of connected two-point moments, $\lexp \de_{1}^{p}
\de_{2}^{q} \rexpc$, which fully characterize the connected two-point
PDF. At large scales it follows that (Bernardeau 1996)

\begin{equation}
    \lexp \de_{1}^{p} \de_{2}^{q} \rexpc = C_{pq}\ 
    \sigma_{L}^{2(p+q-2)}\  \xi_{L} \label{eq:cpqPT},
\end{equation}

where $C_{pq}$ are numbers that depend weakly on scale through
derivatives of the linear variance and smoothed two-point function
with respect to scale ($L$ denotes linear quantities).  Since all
connected moments are proportional to $\xi_{L}$, the connected
two-point PDF should also be proportional to $\xi_{L}$, as in
Eq.~(\ref{eq:Pfac}), where $\xi$ is replaced by $\xi_{L}$.  In
addition, it can be shown that $C_{pq}=C_{p1}C_{q1}$, which leads
directly to the factorization $b(\de_{1}) b(\de_{2})$, with $b(\de)$
calculable from tree-level PT. As loop corrections are included, a
subset of loop diagrams will lead to non-linear corrections to
$\xi_{L}$ and $\sigma_{L}^{2}$ in Eq.~(\ref{eq:cpqPT}), making
$\xi_{L} \rightarrow \xi$ and $\sigma^{2}_{L} \rightarrow \sigma^{2}$
in Eq.~(\ref{eq:cpqPT}).  In addition, there will be new links in the
diagrams leading to higher powers of $\xi_{L}$ and $\sigma^{2}_{L}$;
however, at a given order in PT since $\xi_{L} \ll \sigma^{2}_{L}$ the
latter are much more important than the former.  As a result of this,
we expect that as non-linear scales are probed, as long as $\xi \ll
\sigma^{2}$, Eq.~(\ref{eq:cpqPT}) becomes

\begin{equation}
    \lexp \de_{1}^{p} \de_{2}^{q} \rexpc \sim  C_{pq}(\sigma)\ 
    \sigma^{2(p+q-2)}\  \xi \label{eq:cpq},
\end{equation}

where we have absorbed additional $\sigma$ dependence into $C_{pq}$. 
Note that this also respects the scaling expected in the highly
non-linear regime, as long as $C_{pq}(\sigma) \rightarrow {\rm
const.}$ as $\sigma \rightarrow \infty$.  Recent studies of the
behavior of $\lexp \de_{1}^{p} \de_{2}^{q} \rexpc$ as a function of
$\sigma$ and $\xi$ in numerical simulation support the above arguments
for $\xi \ll \sigma^{2}$ (Gazta\~naga, Fosalba \& Croft 2001). 
However, one must keep in mind that they studied CDM models at $z=0$
rather than $z \sim 3$ as concerns us here\footnote{In any case, even
if there were higher-order corrections in $\xi$ in Eq.~(\ref{eq:cpq}),
still what enters is the {\em non-linear} $\xi$, as shown in
Scoccimarro (2001).  Therefore, our arguments about the difficulties
of determining the initial shape of the dark matter power spectrum
from the forest still apply.}.

%

Now let's consider the effect of redshift distortions.  In this case,
the redshift-space position ${\bf s}= \x -f u_{z}(\x)$, where $f
\approx \Omega_{m}^{0.6} \approx 1$ at the redshifts of interest and
$u_{z}$ is the line of sight velocity, normalized by the Hubble
constant so that in linear PT, $\nabla \cdot {\bf u}= \delta$.  The
transformation between optical depth in real and redshift space reads
$\tau_{s} ds = \tau dz$ (e.g. Hui et al. 1997b ; Gazta\~naga \& Croft
1999) with $ds= J dz$ and $J=|1-f \nabla_{z} u_{z}|$.  This leads to a 
{\em three-dimensional} flux power spectrum (following the calculation 
in Scoccimarro, Couchman \& Frieman 1999), 

\begin{equation}
    P_{s}^{F}(\k) = \int d^{3}r \ e^{i \k \cdot {\bf r}} \lexp 
    e^{i k_{z} f \Delta u_{z}} J_{1}J_{2} e^{-\tau_{1}/J_{1}} 
    e^{-\tau_{2}/J_{2}} \rexp 
    \label{eq:PkFs},
\end{equation}

where ${\bf r} \equiv \x_{1}-\x_{2}$ and $\Delta u_{z} \equiv
u_{z}(\x_{1})-u_{z}(\x_{2}) $.  At large scales, one is allowed to
expand all the exponential factors and use $\de,\nabla_{z} u_{z} \ll
1$, in which case $e^{i k_{z} f \Delta u_{z}} J_{1}J_{2} \approx 1 +
{\cal O}(\nabla_{z} u_{z})^{2}$, and the remaining exponentials give
$P_{s}^{F}(k) =(A\nu)^{2} (1 + \beta \mu^{2})^{2} P_{L}(k)$ (Hui
1999), where we assumed $\tau \equiv A (1+\de)^{\nu}$ and $\mu \equiv
k_{z}/k$, $\beta \equiv f/\nu$.  This is analogous to the standard
result for the galaxy redshift-space power spectrum in the linear
regime (Kaiser 1987).

Non-linear corrections break the relation $\theta \equiv \nabla \cdot
{\bf u}= \de$, valid in linear PT. In particular, a similar
calculation to that in Scoccimarro \& Frieman (1996) for the density
field shows that the critical index for the velocity divergence power
spectrum is $n_{c}^{\theta}\approx -2$.  Therefore, for the models
under consideration with $n\approx -2$ one has $\theta= \nabla \cdot
{\bf u} \sim \de_{L} < \de$; velocity fields are much more ``linear''
than the density field.  We see from Eq.~(\ref{eq:PkFs}) that at small
scales the $J$ factors should suppress the regions undergoing
turnaround, where $J \sim 0$.  These regions, however, correspond to
$1= \nabla_{z} u_{z} \sim \theta /3 \sim \delta_{L}/3$, thus
$\delta_{L} \sim 3$, which corresponds to very high values of $\delta$
and were suppressed already in real space by the exponential
dependence on density.  On the other hand, the effects of the pairwise
velocity along the line of sight encoded in $e^{i k_{z} f \Delta
u_{z}}$ plays a role, at high enough $k_{z}$ oscillations of this
factor damp power along the line of sight leading to a negative
quadrupole to monopole ratio, as in the case of the density field. 
However, since velocity fields should also display critical behavior
as density fields, redshift space corrections should become very
similar for models with different initial spectra.  Therefore, we
believe that the effects of velocities do not affect the conclusions
above.

\section{conclusion}
\label{conclusions}

We have shown that at $z\sim 3$ the constraints on the primordial
power spectrum of density perturbations that can be obtained from the
Lyman alpha forest flux power spectrum are significantly affected by
non-linear corrections to the evolution of the density perturbations.
Because the primordial index of the power spectrum is so negative on
the scales of interest non-linear corrections are much larger than
naively expected. These corrections drive the power spectrum towards a
critical slope, $n\approx -1.4$ irrespective of the primordial spectral
index. This makes the determination of the primordial spectral index
from measurements of the flux power spectrum particularly difficult.

If the objective is to constraint the slope of the primordial power
spectrum, over a wide range of models in the relevant part of
parameter space non-linear corrections are very large. 
Figure~\ref{k20fig} shows that all the scales probed by the forest are
significantly affected by the non-linear corrections for some shapes
of the power spectrum.  Thus, when trying to constrain cosmological
parameters, the flux power spectrum cannot be used as a probe of the
{\it linear} power spectrum of fluctuations at $z\sim 3$ except at the
largest scales.  Figure~\ref{sgneff} illustrates how this
``degeneracy'' manifests itself when determining $n_p$, and
figure~\ref{psfig} shows an example of two models with originally very
different $n_p$ that become almost indistinguishable due to non-linear
evolution.  This problem is made worse by the sensitivity of the bias
to $\bar F$.

There are several ways to increase the sensitivity of the forest to the
shape of the primordial power spectrum.  First future improvements in
measurements of the equation of state might help reduce the error on
the shape. Second, the Sloan Digital Sky Survey (SDSS) has the
capability of extending flux power spectrum measurements out to larger
(and therefore more linear) scales, which will help towards tightening
constraints on the shape of the power spectrum.

\vskip 0.5cm

Acknowledgments: We thank Joop Schaye for useful comments and Enrique
Gazta\~naga for discussions about the behavior of $C_{pq}$ in
numerical simulations.  We thank Patrick MacDonald for pointing out an
error in the first manuscript. MZ is supported by David and Lucille Packard
Foundation Fellowship for Science and Engeneering and NSF grant
AST-0098506.  MZ and RS are supported by NSF grant PHY-0116590.  LH is
supported by an Outstanding Junior Investigator Award from the DOE and
by NSF grant AST-0098437.


\begin{thebibliography}{ZZZZZZZZZZZ1999}
%
%

\vskip 1cm

\bibitem[Alcock \& Paczynski 1979]{alcpac79}
Alcock C., Paczynski B. 1979, Nature, 281, 358 

\bibitem[Bernardeau 1996]{Bernardeau96}
Bernardeau F. 1996, \aap, 312, 11

\bibitem[Bernardeau \& Schaeffer 1999]{BeSc99}
Bernardeau F., Schaeffer,ÊR. 1999, \aap, 349, 697

\bibitem[Bi \& Davidsen 1997]{BD97}
Bi H. G., Davidsen A. F. 1997, \apj, 479, 523

\bibitem[Bryan et al. 1999]{greg99}
Bryan G., Machacek M., Anninos P., Norman M. L. 1999,
\apj, 517, 13

\bibitem[Cen et al. 1994]{chen94}
Cen R., Miralda-Escude J., Ostriker J. P., Rauch M. 1994, 
\apj, 437, L9 

\bibitem[Choudhury et al. 2001]{paddy01}
 Choudhury, T. R., Srianand, R., Padmanabhan, T. 2001
\apj, 559, 29

\bibitem[Croft et al. 1997]{croft97}
Croft R. A. C., Weinberg D. H., Hernquist L., Katz N. 1997, 
\apj, 488, 532

\bibitem[Croft et al. 1998]{croft98}
Croft R. A. C., Weinberg D. H., Katz N., Hernquist L. 1998, 
\apj, 495, 44

\bibitem[Croft et al. 1999]{croft99}
Croft R. A. C., Weinberg D. H., Pettini M., Hernquist L., Katz N. 1999, 
\apj, 520, 1

\bibitem[Croft et al. 2000]{croft00}
Croft R. A. C., Weinberg D. H., Bolte M., Burles S., Hernquist L., Katz N., 
Kirkman D., Tytler D., 2000, astro-ph/0012324

\bibitem[Frieman \& Gazta\~naga 1999]{FrGa99}
Frieman J. A.,  Gazta\~naga E. 1999, \apj, 521, L83

\bibitem[Fry\& Gazta\~naga 1993]{FrGa93}
Fry,ÊJ.ÊN., Gazta\~naga,ÊE. 1993, \apj, 413, 447 


\bibitem[Gazta\~naga 1994]{Gaztanaga94}
Gazta\~naga E. 1994, \mnras, 268, 913

\bibitem[Gaztanaga \& Croft 1999]{gaztacroft}
Gaztanaga E., Croft R. A. C. 1999
\mnras 309, 885

\bibitem[Gaztanaga et al 2001]{gafocr01}
Gaztanaga E., Fosalba P., Croft R. A. C. 2001, astro-ph/0107523

\bibitem[Gnedin \& Hui 1996]{GH96}
Gnedin N. Y., Hui L., 1996, \apjl, 472, 73

\bibitem[Gnedin \& Hui 1998]{gh98}
Gnedin N. Y., Hui L., 1998, \mnras, 296, 44

\bibitem[Goroff et al. 1986]{GGRW86}
Goroff M. H.,  Grinstein B.,  Rey S.-J.,  Wise M. B. 1986, 
\apj, 311, 6

\bibitem[Hernquist et al. 1995]{her95}
Hernquist L., Katz N., Weinberg D. H., Miralda-Escude J. 1995,
\apj, 457, L5 

\bibitem[Hui 1999]{hui99}
Hui L. 1999, \apj, 516, 519

\bibitem[Hui \& Gnedin 1997a]{HuiGne97}
Hui L., Gnedin N. Y. 1997a,
\mnras, 292, 27 
 
\bibitem[Hui et al.  1997b]{Hui97b}
Hui L., Gnedin N. Y., Zhang Y. 1997b,
\apj, 486, 599

\bibitem[Hui and Rutledge 1999]{huirut98}
Hui L., Rutledge R. E. 1999, \apj, 517, 541
 
\bibitem[Kaiser 1987]{kaiser}
Kaiser, N. 1987, \mnras, 227, 1

\bibitem[Kim et al. 1997]{kim}
Kim T. S., Hu E. M., Cowie L. L., Songaila, A. 1997, \aj, 114, 1

\bibitem[LJBH96]{lokas}
{\L}okas E.L., Juszkiewicz R., Bouchet F.R., Hivon E. 1996, \apj, 
467, 1

\bibitem[mss]{mss92}
Makino N., Sasaki M., Suto Y. 1992, \prd, 46, 585

\bibitem[McDonald \& Miralda-Escude 1999]{Mcdjordi}
McDonald, P. \& Miralda-Escude, J. 1999, ApJ, 518, 24.

\bibitem[McDonald et al. 1999]{McD99}
McDonald P., Miralda-Escude J., Rauch M., Sargent W. L. W., Barlow T. a. , Cen R.,
Ostriker J. P., preprint astro-ph/9911196

\bibitem[McDonald et al. 2000]{McD00}
McDonald P., Miralda-Escude J., Rauch M., Sargent W. L. W., Barlow T. a. , Cen R.,
Ostriker J. P., preprint astro-ph/0005553

\bibitem[McDonald et al. 2001]{McD01}
McDonald P., preprint astro-ph/0108064

\bibitem[Machacek and Bertschinger 1995]{edbert} Machacek, 
M. and Bertschinger, E. 1995, American Astronomical Society Meeting, 186, 
0207 

\bibitem[Meiksin \& White 2000]{mw00} 
Meiksin A., White M. 2000, submitted to \mnras, astro-ph 0008214

\bibitem[Meiksin, Bryan \& Machacek 2001]{mbm01} 
Meiksin A., Bryan G., \& Machacek M. 2001, astro-ph/0102367

\bibitem[Miralda-Escude et al. 1996]{m96}
Miralda-Escude J., Cen R., Ostriker J. P., Rauch M. 1996, ApJ, 471, 582

\bibitem[Miralda-Escude et al. 2000]{m00}
Miralda-Escude J., Haehnelt M., Rees M., 2000, \apj, 530, 1

\bibitem[Muecket et al. 1996]{mucket96} Muecket J. P.,
Petitjean P., Kates R. E., Riediger R. 1996, \aap, 308, 17
 
\bibitem[Narayanan et al. 2000]{naray00}
Narayanan V. K., Spergel D. N., Dave R., Ma C. P. 2000,
\apj, 543, L103

\bibitem[Nusser 2000]{nusser00} Nusser A. 2000, \mnras, 317, 902

\bibitem[Nusser \& Haehnelt 2000]{nh99}
Nusser A., Haehnelt M. 2000, \mnras, 313, 364

\bibitem[Pichon et al. 2001]{pichon01}
Pichon, C., Vergely, J. L., Rollinde, E., Colombi, S., Petitjean, P. 2001, 
\mnras, in press, astro-ph/0105196

\bibitem[Rauch et al. 1997]{rauch97}
Rauch, M., Miralda-Escude, J., Sargent, W. L. W., 
Barlow, T. A., Weinberg, D. H., Hernquist, L.,
Katz, N., Cen, R. \& Ostriker, J. P. 1997, ApJ, 489, 7

\bibitem[Reisenegger \& Miralda-Escude 1995]{rm95}
Reisenegger A., Miralda-Escude J. 1995, ApJ, 449, 476

\bibitem[Ricotti et al. 2000]{ricotti00}
Ricotti M., Gnedin N. Y., Shull M. 2000, \apj, 534, 41

\bibitem[Schaye et al. 1999]{joop99}
Schaye J., Theuns T., Leonard A., Efstathiou G. 1999,
\mnras, 310, 57

\bibitem[Schaye et al. 2000]{joop00} Schaye J., Theuns T., Rauch M.,
Efstathiou G., Sargent W. 2000, \mnras, 318, 817

\bibitem[Schaye 2001]{joop01} Schaye J., astro-ph/0104272

\bibitem[Scherrer \& Weinberg 1998]{sw98} Scherrer, R. J., Weinberg, D. H. 
1998, \apj 504, 607

\bibitem[Scoccimarro \& Frieman 1996]{ScFr96}
Scoccimarro R.,  Frieman J. A. 1996, \apj, 473, 620

\bibitem[Scoccimarro, Couchman \& Frieman 1999]{SCF}
Scoccimarro R.,ÊCouchmanÊH.ÊM.ÊP., Frieman,ÊJ. A. 1999, \apj, 517, 531

\bibitem[Scoccimarro 2001]{Sco01}
Scoccimarro R. 2001, Acad. Sci. N.Y., 927, 13, astro-ph/0008277

\bibitem[Theuns et al. 1998]{th98}
Theuns T., Leonard A., Efstathiou G., Pearce F. R.,
Thomas P. A. 1998,
\mnras, 301, 478

\bibitem[Theuns et al. 2000]{th00}
Theuns T., Schaye J., Haehnelt M., \mnras, 315, 600

\bibitem[Theuns et al. 2001]{th01}
Theuns T., Zaroubi S., Tae-Sun Kim, Tzanavaris P., Carswell R., astro-ph/0110600

\bibitem[Tegmark \& Zaldarriaga 2000]{tz00}
Tegmark M., Zaldarriaga M., astro-ph/0002091 

\bibitem[Viel et al. 2001]{viel01}
 Viel, M., Matarrese, S.,  Mo, H. J., Haehnelt, M. G., Theuns, T.
2001, \mnras, in press, astro-ph/0105233

\bibitem[Wadsley \& Bond 1996]{wb96}
Wadsley J. W., Bond J. R. 1996, Proceeding of the 12th Kingston
Conference, eds. Clarke D., West M., PASP, astro-ph/9612148

\bibitem[Wang, Tegmark \& Zaldarriaga 2000]{wtz01}
Wang X., Tegmark M., Zaldarriaga M., astro-ph/0105091 

\bibitem[White \& Croft 2000]{wc2000}
White M., Croft R. A. C. 2000, \apj, 539, 497

\bibitem[Zhang et al. 1995]{zhang95}
Zhang Y., Anninos P., Norman M. L. 1995, 
\apj, 453, L57 

\bibitem[Zaldarriaga et al 2001]{mz00}
Zaldarriaga M., Hui L., Tegmark M. 2001,  \apj, 557,  519

\bibitem[Zaldarriaga et al 2001b]{msh00}
Zaldarriaga M., Seljak U., \& Hui L. 2001b, \apj, 551, 48

\bibitem[Zaldarriaga 2001]{mz01}
Zaldarriaga M., astro-ph 0102205

\end{thebibliography}
\end{document}